\documentstyle[epsfig,psfig]{texas}

\newcommand{\lst}{L_{\ast}}
\newcommand{\msun}{M_{\odot}}
\newcommand{\phist}{\phi_{\ast}}
\newcommand{\lel}{L_{e,1.4}}
\newcommand{\dlogL}{d\log_{10}L}

\newcommand{\cm}{\unit{cm}}
\newcommand{\kmsmpc}{\unit{km\,s^{-1}Mpc^{-1}}} 
\newcommand{\mujy}{\unit{\mu Jy}}
\newcommand{\mum}{\unit{\mu m}}
\newcommand{\myrmpc}{\unit{\frac{\msun}{yr Mpc^3}}}

\newcommand{\ho}{$H_0$}

\newcommand{\unit}[1]{\ifmmode\,{\rm #1}\else$\,{\rm #1}$\fi}
\newcommand{\whzmpc}{\unit{\frac{W}{Hz Mpc^3}}}
\newcommand{\whz}{\unit{\frac{W}{Hz}}}

\begin{document}

\title{FAINT RADIO SOURCES \\ AND STAR FORMATION HISTORY \\
}
\author{Deborah B. Haarsma$^{1}$, R. Bruce Partridge$^{1}$,
	Ian Waddington$^{2}$, and Rogier A. Windhorst$^{2}$  \\
	{\bf For inclusion in the CD-ROM Proceedings of the \\
	19th Texas Symposium on Relativistic Astrophysics, \\
	held December 1998}
}	

\address{(1) Haverford College\\
Haverford, Pennsylvania 19041 USA\\
{\rm Email: dhaarsma@haverford.edu\\}
}

\address{(2) Arizona State University\\
Department of Physics and Astronomy\\
Tempe, Arizona 85287-1504 USA\\
}

\begin{abstract}

Faint extragalactic radio sources provide important information about
the global history of star formation.  Sensitive radio observations of
the Hubble Deep Field and other fields have found that sub-mJy radio
sources are predominantly associated with star formation activity
rather than AGN.  Radio observations of star forming galaxies have the
advantage of being independent of extinction by dust.  We use
the FIR-radio correlation to compare the radio and FIR backgrounds,
and make several conclusions about the star forming galaxies producing
the FIR background.  We then use the redshift distribution of faint
radio sources to determine the evolution of the radio luminosity
function, and thus estimate the star formation density as a function
of redshift.

\end{abstract}

\section{Introduction}
 
Faint radio sources provide important information about global star
formation history.  Sensitive radio observations of the Hubble Deep
Field (HDF) (Richards~\etal 1998, \cite{richards98a}) and other fields
well-studied at optical wavelengths (Windhorst~\etal 1995, 
\cite{windhorst95a}, Fomalont~\etal 1991, \cite{fomalont91a}) have shown
that sub-mJy radio sources are predominantly associated with star
formation activity rather than active galactic nuclei (AGN).  The
radio luminosity of a galaxy is a reliable predictor of the star
formation rate (SFR) for local galaxies (Condon~1992, \cite{condon92a}
Cram~\etal 1998, \cite{cram98a}).  Estimates of star formation based on
radio observations also have the advantage of being independent of
extinction by dust, which has caused much difficulty in the
determination of star formation history from optical data.
 
In section 2, we make use of the tight correlation between radio and
FIR luminosity for star forming galaxies to compare
the FIR and radio backgrounds and to study the sources producing both.
In section 3, we determine the evolving radio luminosity function from the
observed redshift distribution of faint radio sources, and then
estimate the history of star formation to a redshift of about 3.
Throughout, we assume $\Omega_m=1$, $\Omega_\Lambda=0$, and
\ho=50\kmsmpc.  We define the radio spectral index as
$S_\nu\propto\nu^{-\alpha}$.

\section{FIR vs. Radio Backgrounds}
 
\begin{figure}
\centering
\epsfig{figure=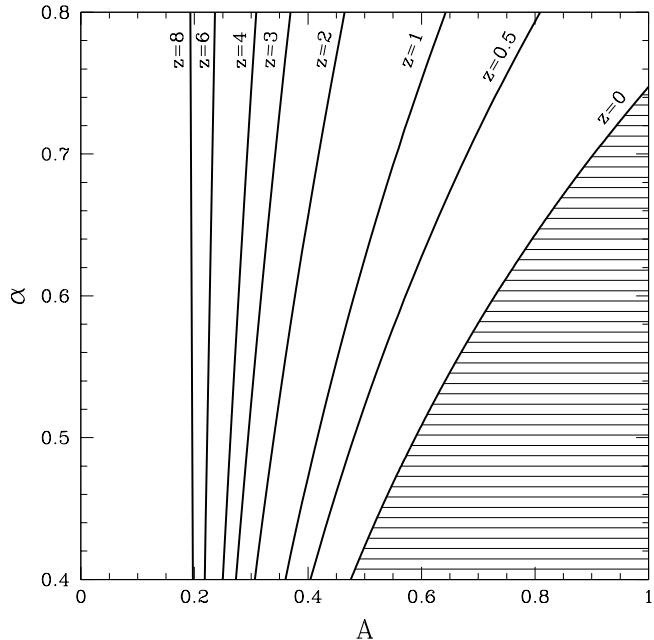, width=9cm}
\caption{ 
Relationship between the radio spectral index $\alpha$, the ratio of
star-formation flux to the total radio background $A$, and the typical
redshift $z$ for the sources making up the FIR background.  
}
\label{fig.Azalpha}
\end{figure}

This section follows our recent paper (Haarsma~\& Partridge 1998,
\cite{haarsma98a}).  The far infrared (FIR) background was recently
detected with DIRBE (Hauser \etal1998, \cite{hauser98a}; Dwek
\etal1998, \cite{dwek98a}), and is most likely the collective emission
of star forming galaxies.  We use the radio-FIR correlation for
individual galaxies (Helou, Soifer, \&~Rowan-Robinson 1985,
\cite{helou85a}) to calculate the radio background associated with the
FIR background, assuming that the bulk of emission is from $z\sim 1$.
We find the radio background associated with the FIR background has a
brightness temperature of $T_{40\cm}=0.31$~K, or $T_{170\cm}\sim15$~K
(scaled using a spectral index of $\alpha=0.7$).  At 170\cm\
(178~MHz), the observed radio background is $T_{170\cm}=30\pm7$~K
(Bridle 1967, \cite{bridle67a}).  This allows us to draw several
conclusions about the faint sources making up the FIR background:
 
\begin{enumerate}
 
\item
The radio emission from these sources makes up about half of the
observed extragalactic radio background.  (The other half is the
summed radio emission of AGN.)
 
\item
Since (i) is in agreement with other radio observations (Condon 1989, 
\cite{condon89a}), the FIR-radio correlation appears to hold even for
the very faint sources making up the FIR background.  This confirms
the assumption that the FIR background between about 140 and 240\mum\
is dominated by star-formation, not AGN activity.
 
\item
By quantitatively comparing the radio and FIR backgrounds, we find
a relationship for the sources contributing to the
background,
\begin{equation}
	A \left(\frac{ 1+z}{8.5}\right)^\alpha = 0.20\pm0.05,
\end{equation}
where $\alpha$ is their radio spectral index, $A$ is the fraction of
the radio background they produce (from (i), $A\sim0.5$), and $z$ is
their mean redshift.  This function is plotted in
Figure~\ref{fig.Azalpha}.  Note that the redshift $z$ is the mean
redshift of the sources dominating the FIR and radio backgrounds,
which is not necessarily the redshift of peak star formation activity
(see \S3).
 
\item
By extrapolating the 3.6\cm\ $\log N-\log S$ curve to fainter flux
densities, we estimate that most of the FIR background is produced by
sources whose 3.6\cm\ flux density is greater than about 1\mujy.  This
lower limit is consistent with other work (Windhorst~\etal 1993, 
\cite{windhorst93a}), but has more interesting observational
consequences.  An RMS sensitivity of 1.5\mujy\ has already been
reached in VLA observations (Partridge~\etal 1997, \cite{partridge97a}).
The $\log N-\log S$ curve indicates that the number density of $S\geq1$\mujy\
sources is about $25/\unit{arcmin^2}$, similar to some model
predictions (Guiderdoni \etal 1998, \cite{guiderdoni98a}).  At this
density, these sources will cause SIRTF to encounter confusion
problems at 160\mum.
 
\end{enumerate}

\section{Radio Star Formation History}
 
In this section, we use the redshift distribution of faint radio
sources to determine the evolution of the radio luminosity function,
and the evolution of the star formation rate density.

\subsection{Data}
Three fields have been observed to microJy sensitivity at centimeter
wavelengths and also have extensive photometric and spectroscopic
data: the Hubble Deep Field (HDF), the Medium Deep Survey (MDS), and
the V15 field.  Table~\ref{tab.data} gives the details of the three
fields and references.  For the first time we have a sample of microJy
radio sources with nearly complete optical identifications and about
50\% complete redshift measurements.  We assume that all sources
detected at these flux levels are star-forming galaxies, since optical
identifications indicate that $\sim$80\% of these radio sources have
spiral or irregular counterparts (Richards \etal 1998,
\cite{richards98a}).  The known quasars (two in the MDS sample, none
in the HDF or V15 samples) were removed. In the flanking fields of the
HDF, we have used the relationship between redshift and K-band
magnitude (Lilly, Longair, \&~Allington-Smith 1985, \cite{lilly85a})
to estimate redshifts for 10 sources without spectroscopic values.
For the remaining sources without redshifts, we arbitrarily selected
redshifts to fill in gaps in the redshift distribution, in order to
illustrate the total number of sources that will ultimately appear on
the plot. Photometric redshifts for these sources are currently being
calculated (Waddington~\& Windhorst, in preparation), and will be
included in future work.

To compare these data to the model, we calculate $n(z)$, the average
number of sources per arcmin$^2$ in each redshift bin.  This requires
a correction for the varying sensitivity across the primary beam of
the radio observations (Katgert, Oort, \& Windhorst 1988,
\cite{katgert88a}; Martin, Partridge, \& Rood 1980, \cite{martin80a}).
For example, a faint source which could only be detected at the center
of the field contributes more to $n(z)$ than a strong source which
could be detected over the entire primary beam area.  The resulting
redshift distributions are plotted in Figures~\ref{fig.modelC.nz} and
\ref{fig.modelS.nz}.

It is interesting how different the MDS and HDF distributions are,
even though the surveys were both performed at 8~GHz with similar flux
limits.  The average source density (including all sources) is 1.26
sources/arcmin$^2$ in the HDF, but 2.63 sources/arcmin$^2$ in the MDS
field (the V15 field at 5~GHz has 0.736 sources/arcmin$^2$).  The
density of sources in the MDS field is over twice that of the HDF
field, possibly due to galaxy clustering.  In the analysis below, we
fit the model to the data in all three fields simultaneously.

\begin{table}[t]
\begin{center}
\begin{tabular}[]{lllllll}
\hline \hline
Field              & Location & Band  & Flux limit  & N  & N$_z$   & Reference \\
\hline
Hubble Deep Field  & 12h+62d  & 8 GHz & 9\mujy      & 29 & 13 (+10)& \cite{richards98a}  \\
Medium Deep Survey & 13h+42d  & 8 GHz & 8.8\mujy    & 19 & 10      & \cite{windhorst95a} \\
V15 field          & 14h+52d  & 5 GHz & 16\mujy     & 35 & 18      & \cite{fomalont91a,hammer95a} \\
\hline
\end{tabular}
\vspace{3mm}
\caption{Summary of three deep radio surveys.  The flux limit of the
complete catalog for each field is approximately 5 times the RMS noise
of the observation, but varies across the field.  N is the
total number of sources above the flux limit, and $N_z$ is the number
of those sources with spectroscopic redshifts.  An additional 10
redshifts were estimated in the HDF from their K-band magnitudes. }
\label{tab.data}
\end{center}
\end{table}

\subsection{Calculations}
In order to determine the star formation history, we must first
determine the evolving radio luminosity function.  We used two
versions of the local 1.4~GHz luminosity function for
star-forming/spiral galaxies.  We define the luminosity function
$\phi(\lel)$ as the number per comoving Mpc$^3$ per $\dlogL$ of
star-forming radio sources with emitted luminosity $\lel$(W/Hz)at
1.4~GHz.  Condon (1989, \cite{condon89a}) uses the following form for
the luminosity function (but different notation),
\begin{eqnarray}
	\log_{10}[\phi(\lel)] & \dlogL = 28.43 + Y  
	- 1.5\log_{10}\lel  \nonumber \\
	& - \left[ B^2 + \frac{1}{W^2} (\log_{10}\lel - X)^2 \right]^{1/2}
	   \dlogL,
\label{eq.lumfunc.C89}
\end{eqnarray}
with the fitted parameters for star-forming galaxies of $Y=2.88$,
$X=22.40$, $W=2/3$, and $B=1.5$.  Serjeant \etal (1998,
\cite{serjeant98a}) use the standard Schechter form,
\begin{equation}
	\phi(\lel)\dlogL = \phist \ln10 
		\left(\frac{\lel}{\lst}\right)^{(1+\alpha_l)}
		\exp\left(-\frac{\lel}{\lst}\right) \dlogL
\label{eq.lumfunc.S98}
\end{equation}
where a factor of $L\ln10$ has been included to convert the function
from d$L$ to $\dlogL$.  Serjeant \etal find fitted parameters of
$\phist=4.9\times10^{-4}\unit{Mpc^{-3}}$,
$\lst=2.8\times10^{22}\unit{W/Hz}$, and $\alpha_l = -1.29$.

To describe the evolution of the luminosity function, we use the
functional form suggested by Condon (1984a, \cite{condon84a}, eq.~24),
a power-law in $(1+z)$ with an exponential cut-off at high redshift.
The luminosity evolves as
\begin{equation}
	f(z) = (1+z)^Q \exp\left[ -\left(\frac{z}{z_q}\right)^q \right],
\label{eq.lumevol}
\end{equation}
and the number density evolves as
\begin{equation}
	g(z) = (1+z)^P \exp\left[ -\left(\frac{z}{z_p}\right)^p \right].
\label{eq.numevol}
\end{equation}
This gives six parameters $\{Q,q,z_q,P,p,z_p\}$ to use in describing
the evolution.  When fitting for the parameters, we constrained the
functions $g(z)$ and $f(z)$ to the physically reasonable ranges of
$1<g(z)<100$, $1<f(z)<100$ for $0<z<3$.
The general expression for the evolving luminosity function
is then (Condon 1984b, \cite{condon84b})
\begin{equation}
	\phi(\lel,z) = g(z) \phi\left( \frac{\lel}{f(z)}, 0 \right).
\label{eq.evolvedlumfunc}
\end{equation}
To use this expression at an arbitrary observing frequency $\nu$ and
redshift $z$, we must convert the observed luminosity $L_{o,\nu}$ to
1.4~GHz and do the K-correction, i.e.
\begin{equation}
	\lel = L_{o,\nu} 
	        \left( \frac{\nu }{ 1.4\unit{GHz}} \right)^\alpha 
	        (1+z)^\alpha
\end{equation}
where $\alpha$ is the radio spectral index, as defined in \S1.  We
have assumed $\alpha=0.4$ for all calculations in \S3 (Windhorst
\etal 1993, \cite{windhorst93a}).


The evolving luminosity function can be used to predict the observed
redshift distribution.  The number of sources per redshift bin $\Delta
z$ that could be detected in a survey of angular area $\Delta\Omega$
and flux limit $S_{lim}$ at frequency $\nu$ is
\begin{equation}
	n(z) = 
	V_c(z,\Delta z,\Delta\Omega) \int_{L'(z)}^{\inf} \phi(\lel,z) \dlogL
\end{equation}
where the lower limit of the integral is
\begin{equation}
	L'(z) = 9.5\times10^{12}\whz
	\left( \frac{S_{lim} }{\mujy         } \right)
	(1+z)^\alpha 
	\left( \frac{\nu     }{1.4\unit{GHz} } \right)^\alpha
	4 \pi 
	\left( \frac{D_L(z)  }{\unit{Mpc}    } \right)^2
\end{equation}
and $D_L$ is the luminosity distance.
The comoving volume in a shell from $z$ to $z + \Delta z$ and angular size
$\Delta\Omega$ is
\begin{eqnarray}
	V_c(z, \Delta z, \Delta\Omega) 
	& = & \int d\Omega \int r^2(z) dr \nonumber \\
	& = & 
              \frac{\Delta\Omega    }{ \unit{ster}                }
	\left(\frac{\unit{ster}     }{ 1.18\times10^7\unit{arcmin^2}}\right)
	\frac{[r^3(z+\Delta z) - r^3(z)]  }{ 3                    }
\label{eq.covol}
\end{eqnarray}
where the comoving distance is 
\begin{equation}
	r(z) = \frac{2c}{H_0} \left( 1 - \frac{1}{\sqrt{1+z}} \right)
\end{equation}
for our assumed cosmology (see \S1).  We have used 
$\Delta \Omega = 1$~arcmin$^2$ for comparison to the data in 
Figures~\ref{fig.modelC.nz} and \ref{fig.modelS.nz}.

The evolving luminosity function also allows us to calculate the star
formation history.  For an individual galaxy, the star formation rate
is directly proportional to its radio luminosity (Condon 1992, \cite{condon92a}):
\begin{equation}
	\unit{SFR} = Q
	\left( \frac{L_\nu /\whz 
		}{5.3\times10^{21} \left(\frac{\nu}{\unit{GHz}}\right)^{-0.8} 
	 	+ 5.5\times10^{20} \left(\frac{\nu}{\unit{GHz}}\right)^{-0.1} } 
	\right)\unit{\frac{\msun}{yr}}
\label{eq.sfrtoLpergal}
\end{equation}
The radio luminosity is primarily due to synchrotron emission from
supernova remnants (the first term in the denominator) plus a small
thermal component (the second term).  Both components are proportional
to the formation rate of high-mass stars which produce supernova
($M>5\msun$), so the factor $Q$ is included to account for the mass of
all stars ($0.1-100\msun$),
\begin{equation}
	Q = \frac{  \int_{0.1\msun}^{100\msun} M \psi(M) dM 
	        }{  \int_{  5\msun}^{100\msun} M \psi(M) dM 
                },		
\end{equation}
where $\psi(M)\propto M^{-x}$ is the initial mass function (IMF).
We have assumed throughout a Salpeter IMF ($x=2.35$), for which $Q=5.5$.
If an upper limit of 125$\msun$ is used, then $Q=5.9$. 

In order to use eq.~\ref{eq.sfrtoLpergal} at high redshift, both
$L_\nu$ and $\nu$ in the equation must be K-corrected to the emitted
luminosity at the emission frequency.  Are there other ways in which
this relation evolves?  The thermal term is much smaller than the
synchrotron term, so evolution in the thermal term will have little
effect.  In the synchrotron term, the dependence on the supernova
environment is weak.  One component that might cause significant
evolution in eq.~\ref{eq.sfrtoLpergal} is an evolving IMF, entering
through the factor $Q$.  In active starbursts, the IMF may be weighted
to high-mass stars (Elmegreen 1998, \cite{elmegreen98a}), which would
result in a smaller value of $Q$.  However, the smallest $Q$ is unity
(when virtually all mass occurs in high-mass stars), so the strongest
decrease due to IMF evolution would be roughly a factor of five.

To determine the star formation rate per comoving volume, we simply
substitute the radio luminosity density for $L_\nu$ in
eq.~\ref{eq.sfrtoLpergal}.  The star formation rate depends on the
emitted (rather than observed) luminosity density.  The luminosity
density emitted at 1.4~GHz can be easily found from the evolving
luminosity function,
\begin{equation}
	\rho_{e,1.4}(z) = 
	\int_{-\inf}^{\inf} \lel \phi(\lel,z) \dlogL.
\end{equation}
Thus the predicted star formation history is
\begin{equation}
	\psi(z) = Q 
		  \left(\frac{\rho_{e,1.4}(z)}{ 4.6\times10^{21} \whzmpc}\right)
		  \myrmpc
\label{eq.sfrthist}
\end{equation}
where 1.4\unit{GHz} is used in the denominator of
eq.~\ref{eq.sfrtoLpergal} (no K-correction is needed because the
luminosity density is the emitted value).

\subsection{Results}

We use the formulation of \S3.2 to determine the star formation
history from the evolving luminosity function.  To determine the
evolution parameters, we compare the model to the observed $n(z)$ for
the three surveys.  We immediately found that pure luminosity
evolution [$f(z)=(1+z)^3$ and $g(z)=1$], as often suggested in the
literature, is a poor fit for the faint star-forming galaxy population
(the predicted $n(z)$ is too small and has a very long high-redshift
tail).  The model fit of Condon (1984a, \cite{condon84a}), $\{Q=3.5,
P=1.75, p=1.8, z_p=1 \}$ with no exponential cut off in luminosity
evolution, is much better (more reasonable redshift dependence, but
$n(z)$ is still too low).  To improve on these models, we adjust the
evolution parameters $\{Q, q, z_q, P, p, z_p\}$ to improve the model
fit to the $n(z)$ data, using a downhill simplex algorithm (Press
\etal 1992, \cite{recipes2}) to find the global $\chi^2$ minimum.  We
performed this fit using both the Condon (1989, \cite{condon89a})
luminosity function (Model~C, see eq.~\ref{eq.lumfunc.C89}) and the
Serjeant \etal (1998, \cite{serjeant98a}) luminosity function
(Model~S, see eq.~\ref{eq.lumfunc.S98}).

\begin{figure}
\centering
\epsfig{figure=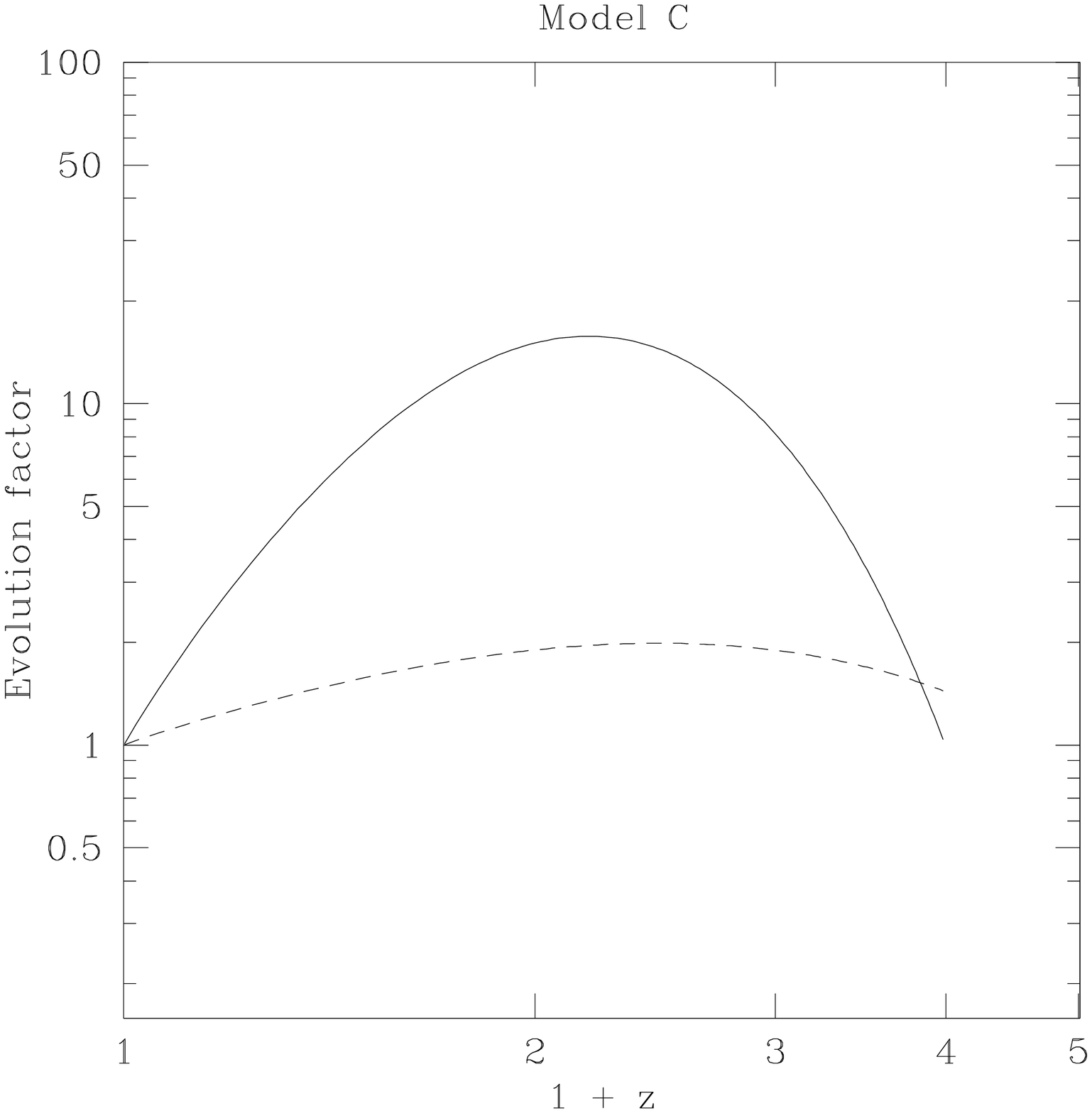, width=0.5\textwidth}
\caption{Evolution functions for Model~C.  The solid line is $f(z)$
(luminosity evolution), and the dashed line is $g(z)$ (number density
evolution).  }
\label{fig.modelC.evolfactor}
\end{figure}

\begin{figure}
\centering
\epsfig{figure=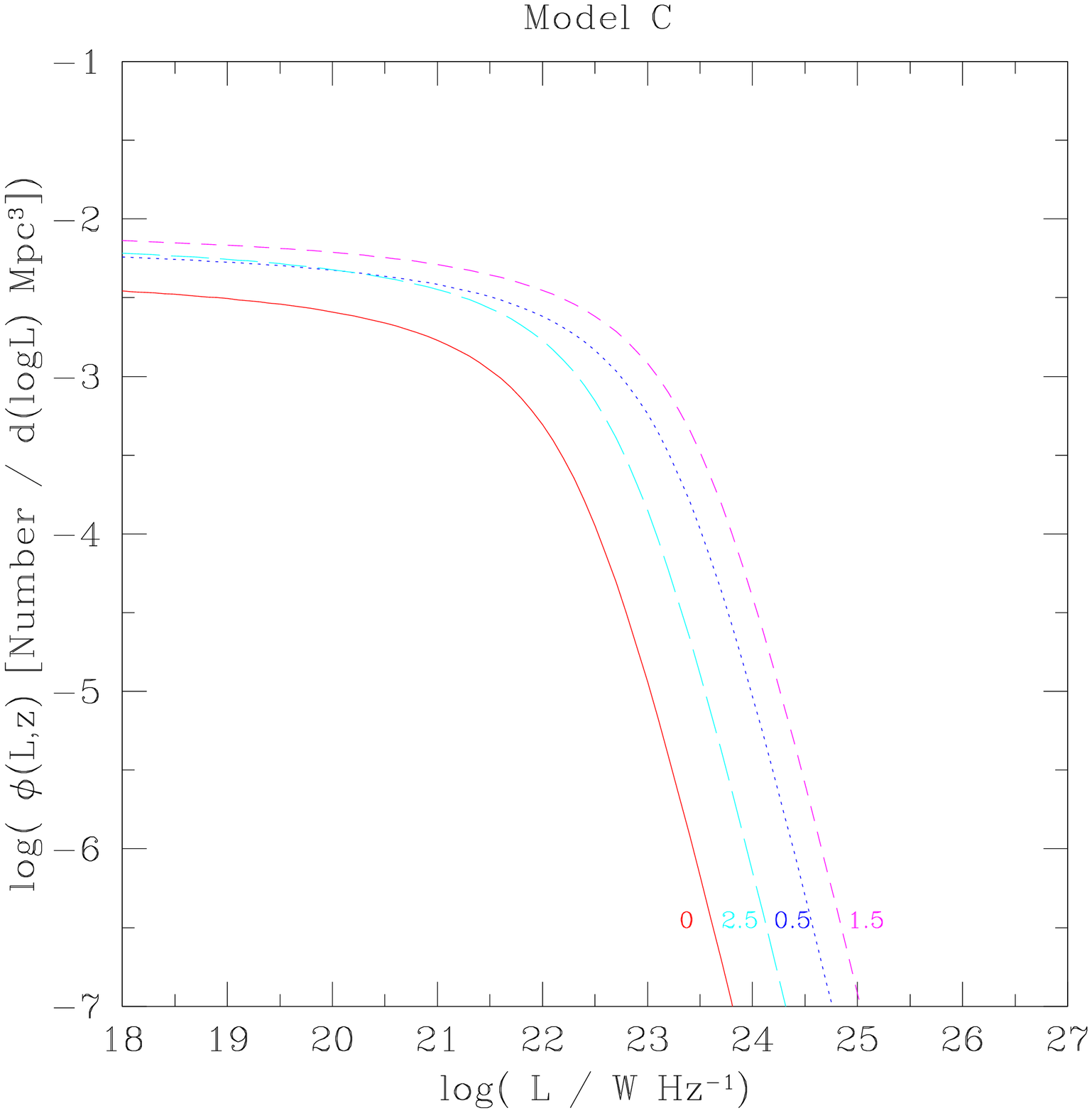, width=0.5\textwidth}
\caption{Evolving luminosity function for Model~C.  The labels indicate
the redshift for each curve, for $z=0,$ 0.5, 1.5, and 2.5.}
\label{fig.modelC.lumfuncevolve}
\end{figure}

\begin{figure}
\centering
\epsfig{figure=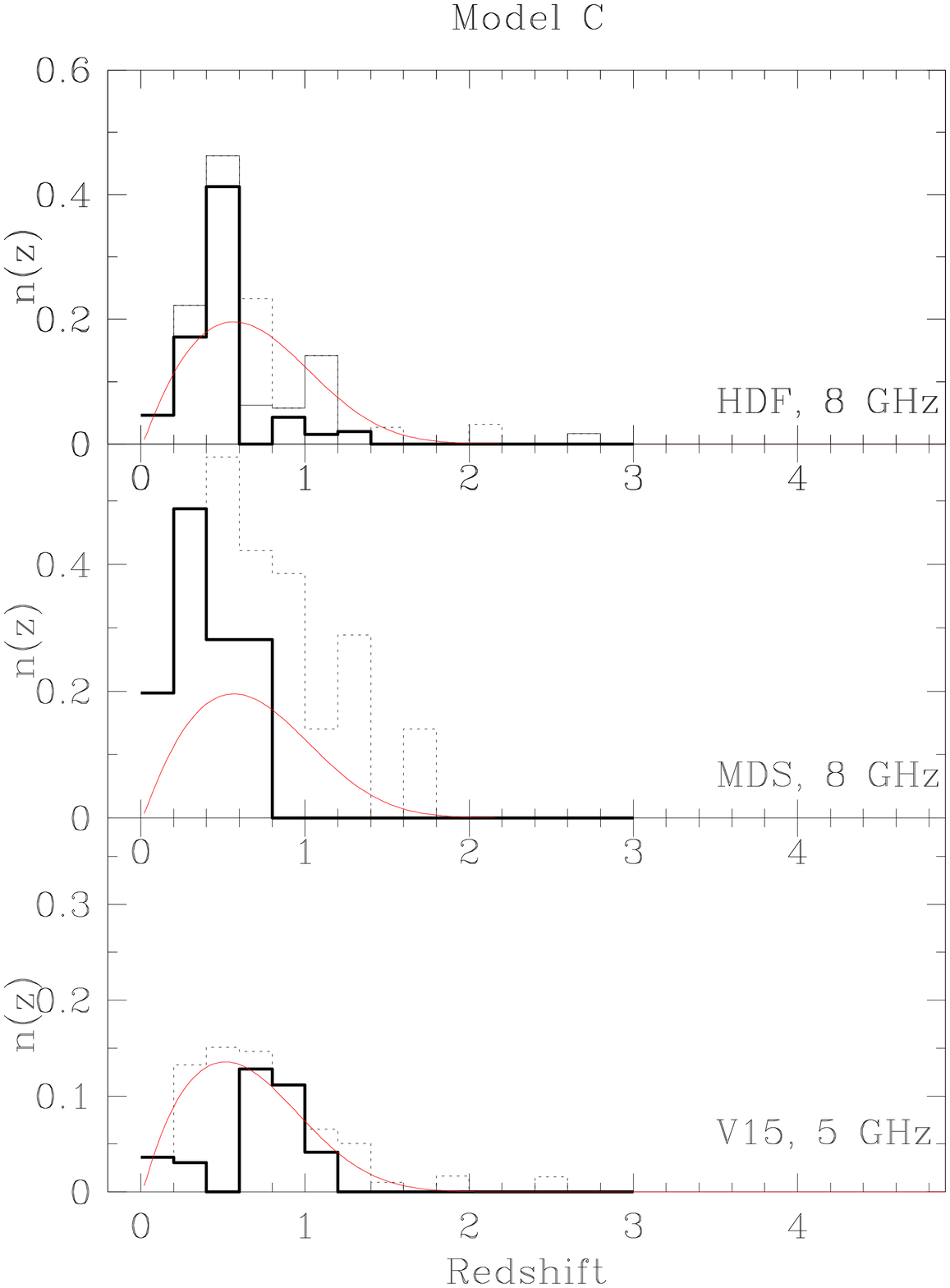, width=0.5\textwidth}
\caption{Redshift distribution $n(z)$, number per arcmin$^2$ per
redshift bin.  The curve shows the prediction of Model~C.  The data
histograms indicate spectroscopic redshifts [heavy line],
spectroscopic redshifts plus estimates from K magnitudes [thin line
histogram], and all sources (including arbitrary redshifts for
remaining sources) [dotted line].  }
\label{fig.modelC.nz}
\end{figure}

\begin{figure}
\centering
\epsfig{figure=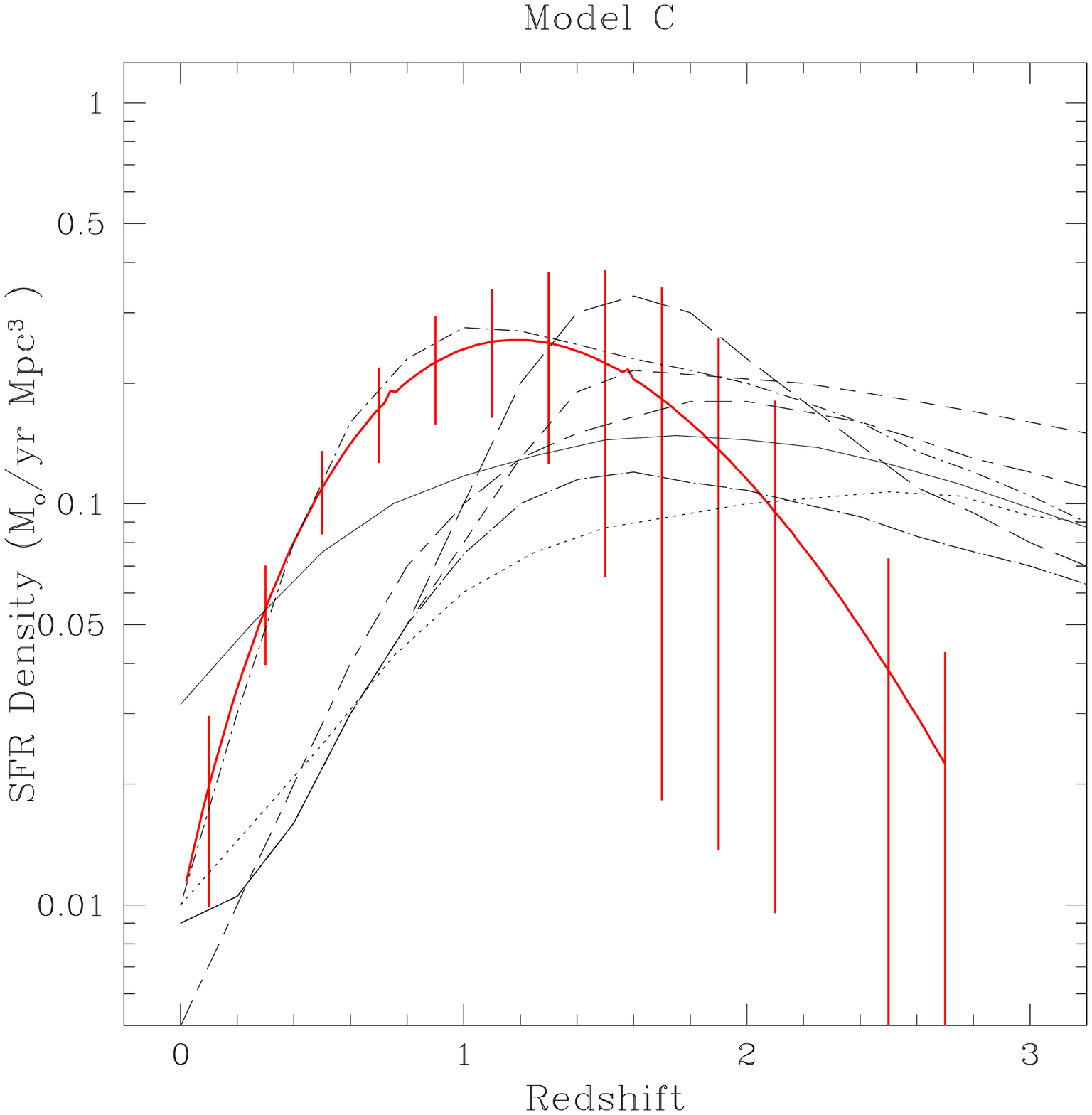, width=0.5\textwidth}
\caption{Star formation history. The heavy curve is the prediction of
Model~C, with vertical lines indicating the uncertainty.  The
remaining lines indicate star formation histories predicted by several
other models: solid line (Baugh~\etal1998, [22]), dotted
line (Guiderdoni~\etal1999 [26]), other broken lines
(Dwek~\etal1998, [8], figure 3).  
}
\label{fig.modelC.sfr}
\end{figure}

In Model~C we use the luminosity function of Condon (1989,
\cite{condon89a}).  The fitted evolution parameters were $\{ Q=7.6,
q=1.3, z_q=0.48, P=1.6, p=1.2, z_p=1.8 \}$. The resulting evolution
factors $f(z)$ and $g(z)$ are plotted in
Figure~\ref{fig.modelC.evolfactor} and the resulting luminosity
function is shown in Figure~\ref{fig.modelC.lumfuncevolve}.  Although
the term $(1+z)^{7.6}$ seems extreme, when combined with the
exponential cut-off the luminosity evolution $f(z)$ is reasonable.
The fit to the the redshift distribution is shown in
Figure~\ref{fig.modelC.nz}.  The fit significantly underestimates the
total number of sources in the MDS field, but only slightly
underestimates the other two fields. The V15 survey has the largest
total number of sources and thus has the most weight during fitting,
so the result is a better fit for V15 than the other fields.  Finally,
Figure~\ref{fig.modelC.sfr} shows our predicted star formation history
(heavy line) along with model predictions from several others (thin
lines).  The vertical lines indicate the $1/\sqrt{N}$ uncertainty,
where $N$ is the sum of galaxies at that redshift from the three
surveys.  The Model~C prediction is in good agreement with other
models at low redshift (the curve follows closely the prediction of
Pei \&~Fall 1995, \cite{pei95a}, as plotted in Dwek \etal 1998,
\cite{dwek98a}, figure 3), which is impressive given that no free
parameters were adjusted to fit the $z=0$ value.  The predicted star
formation history peaks around a redshift of 1, and falls off more
quickly than other models at high redshift.

\begin{figure}
\centering
\epsfig{figure=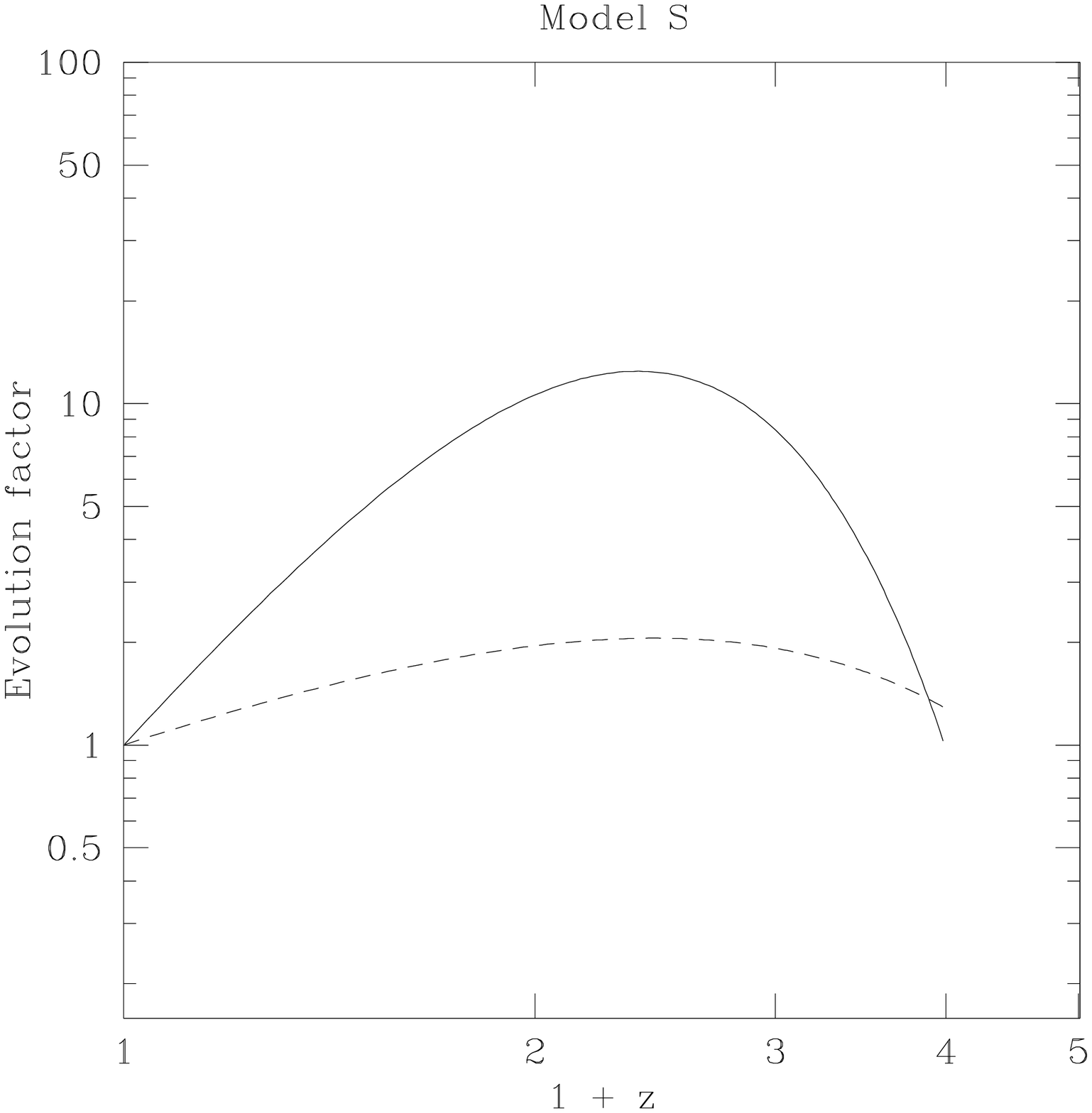, width=0.5\textwidth}
\caption{Evolution functions for Model~S.  The solid line is $f(z)$
(luminosity evolution), and the dashed line is $g(z)$ (number density
evolution).  }
\label{fig.modelS.evolfactor}
\end{figure}

\begin{figure}
\centering
\epsfig{figure=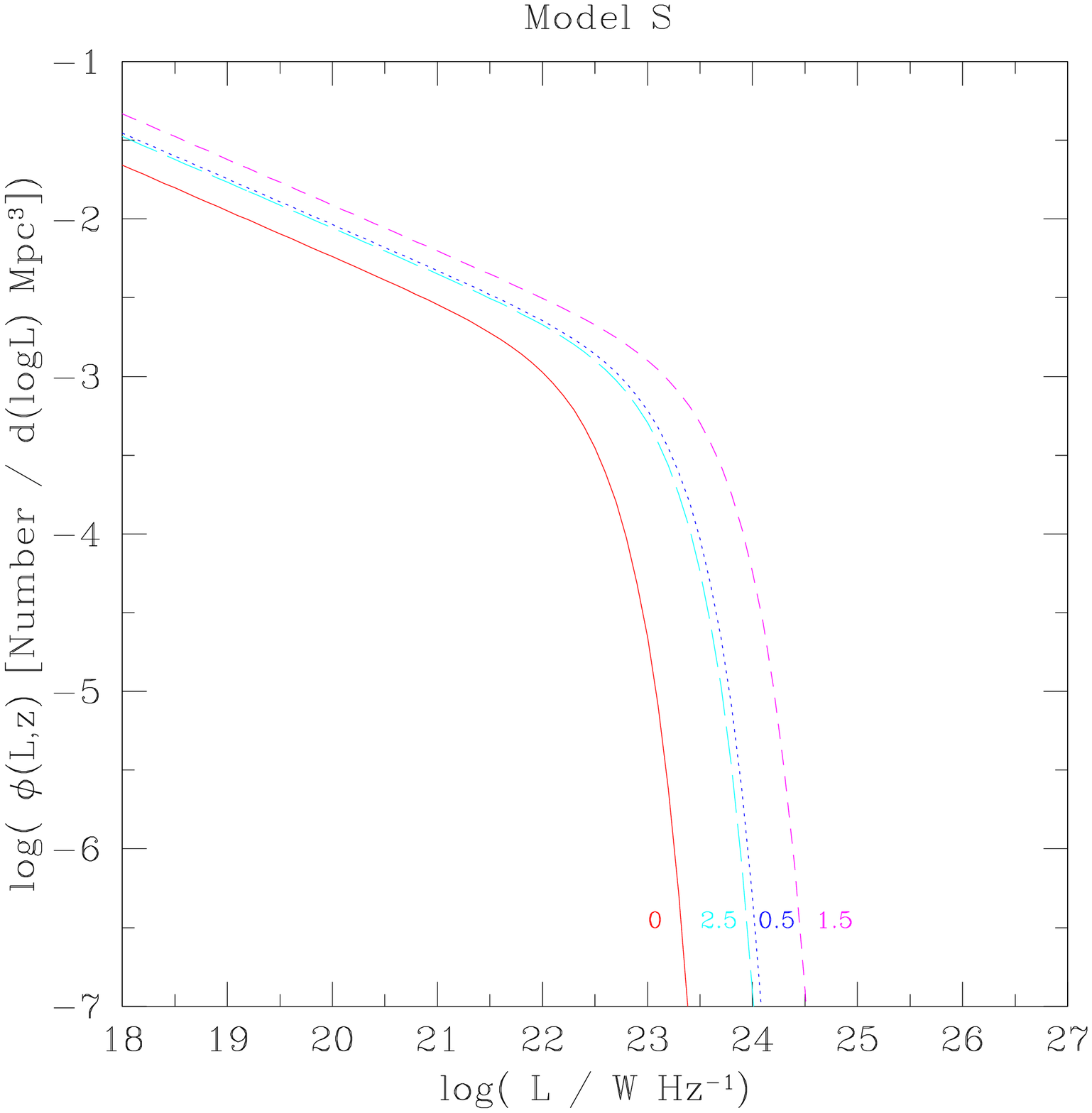, width=0.5\textwidth}
\caption{Evolving luminosity function for Model~S.  The labels indicate
the redshift for each curve, for $z=0,$ 0.5, 1.5, and 2.5.}
\label{fig.modelS.lumfuncevolve}
\end{figure}

\begin{figure}
\centering
\epsfig{figure=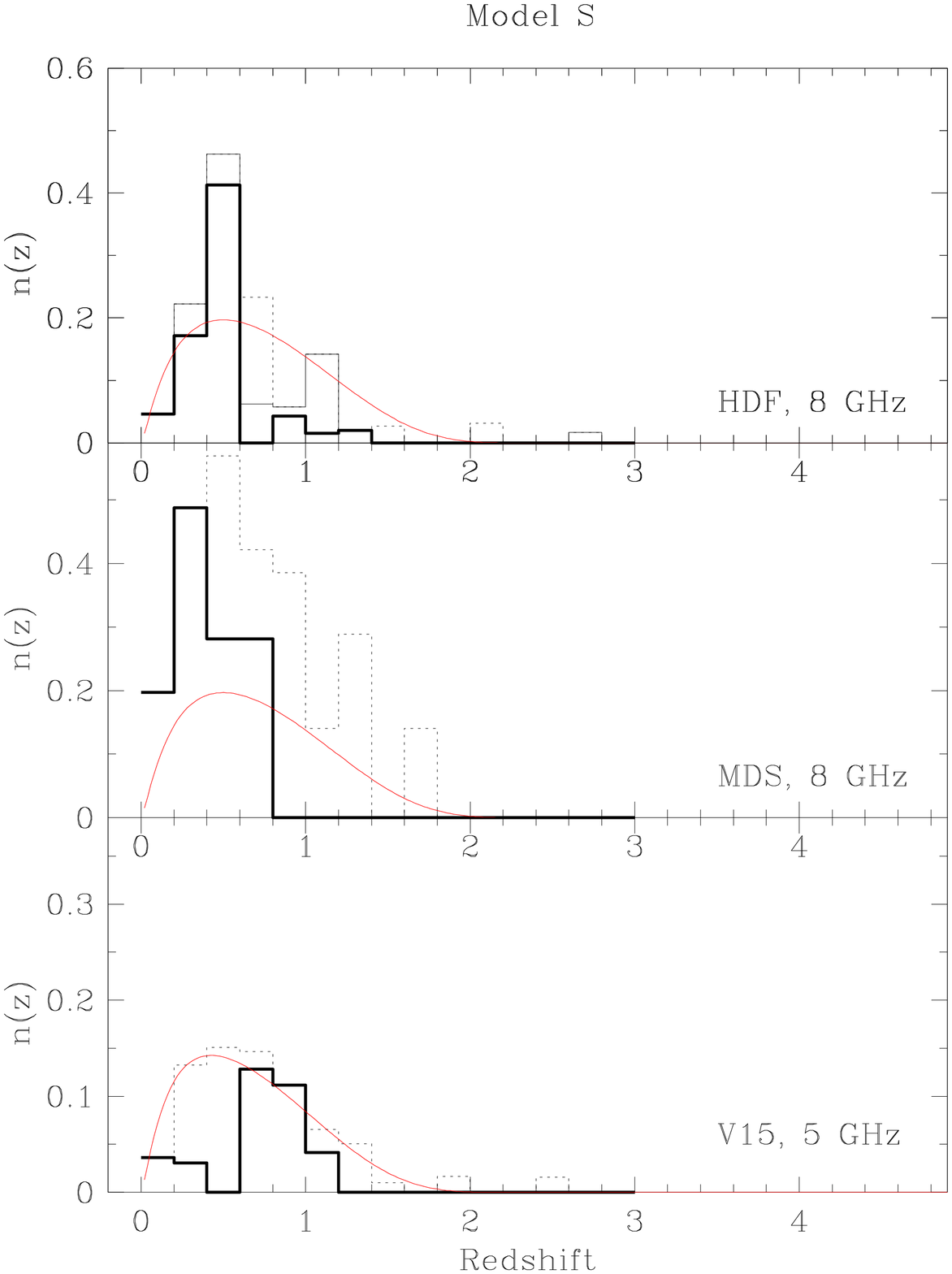, width=0.5\textwidth}
\caption{Redshift distribution $n(z)$, number per arcmin$^2$ per
redshift bin.  The curve shows the prediction of Model~S.  The data
histograms indicate spectroscopic redshifts [heavy line],
spectroscopic redshifts plus estimates from K magnitudes [thin line
histogram], and all sources (including arbitrary redshifts for
remaining sources) [dotted line].  }
\label{fig.modelS.nz}
\end{figure}

\begin{figure}
\centering
\epsfig{figure=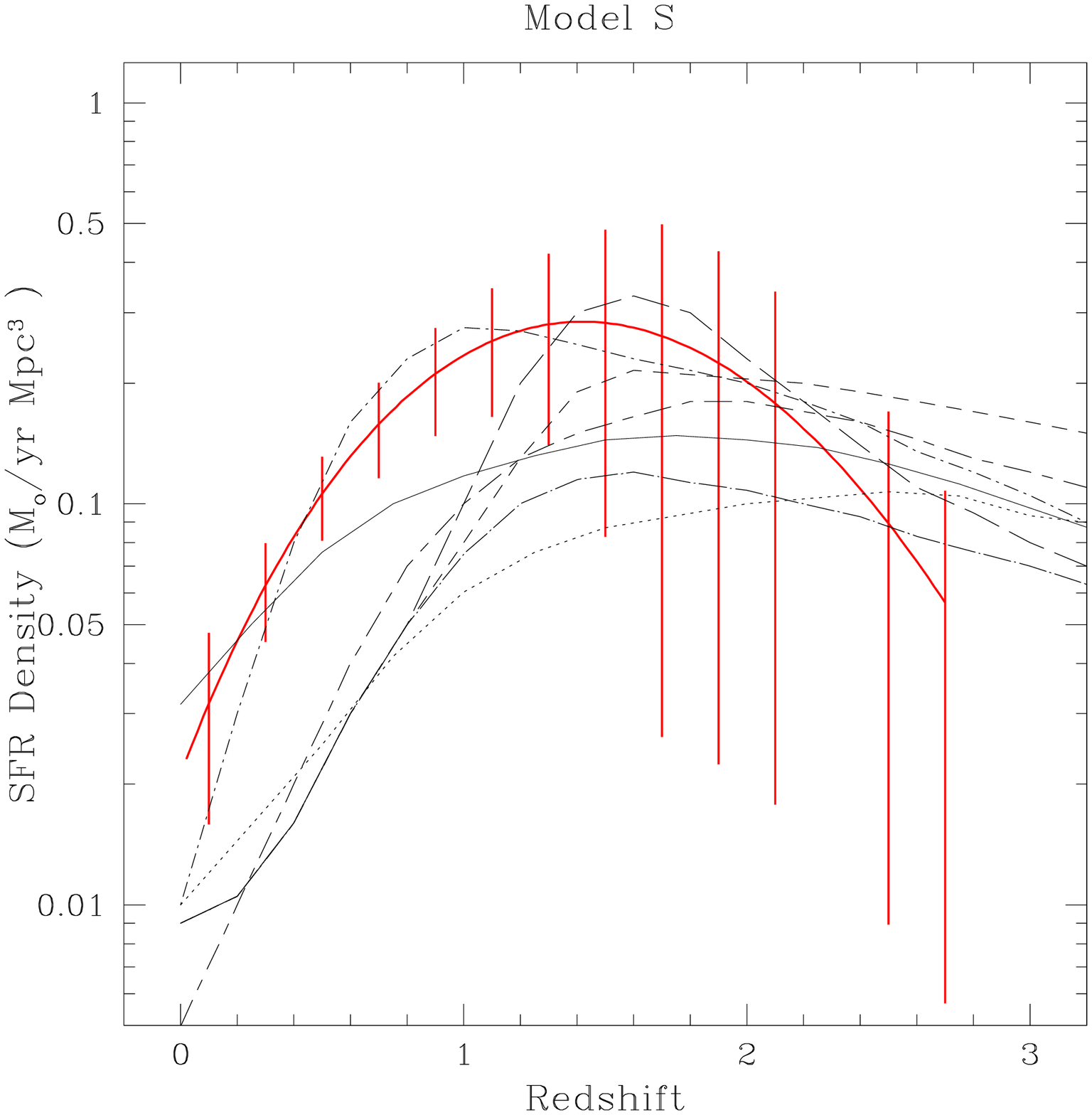, width=0.5\textwidth}
\caption{Star formation history. The heavy curve is the prediction of
Model~S, with vertical lines indicating the uncertainty.  The
remaining lines indicate star formation histories predicted by several
other models: solid line (Baugh~\etal1998, [22]), dotted
line (Guiderdoni~\etal1999, [26]), other broken lines
(Dwek~\etal1998, [8], figure 3).  
}
\label{fig.modelS.sfr}
\end{figure}

In Model~S we use the luminosity function of Serjeant \etal (1998,
\cite{serjeant98a}) (see eq.~\ref{eq.lumfunc.S98} above).  The fitted
evolution parameters were $\{ Q=4.3, q=2.1, z_q=1.3, P=1.3, p=1.7,
z_p=2.3 \}$. The resulting evolution factors $f(z)$ and $g(z)$ are
plotted in Figure~\ref{fig.modelS.evolfactor} and the resulting
luminosity function is shown in Figure~\ref{fig.modelS.lumfuncevolve}.
Despite very different individual parameters ($Q=4.3$ vs. $Q=7.6$),
the two fits have similar functions $f(z)$ and $g(z)$.  The predicted
redshift distribution (Figure~\ref{fig.modelS.nz}) is peaked at a
slightly lower redshift and has a slightly longer tail then Model~C.
The predicted star formation history (Figure~\ref{fig.modelS.sfr}) has
a larger local value than Model~C, but still less than that predicted
by Baugh \etal (1998, \cite{baugh98a}) (thin solid line).  The peak is around a redshift
of 1.4, and falls off less rapidly than Model~C at high redshift.

\subsection{Discussion}

The star formation histories predicted by Model~C and Model~S both
fall off more quickly at high redshift than model predictions by
others.  However, we are considering several refinements to our model
that might modify this result.  We are currently determining
additional photometric redshifts (Waddington \& Windhorst, in
preparation), which will make the modeling more reliable particularly
at high redshift.  The predicted shape of the star formation history
is limited by the functional form we chose for evolution
(eq.~\ref{eq.lumevol} and \ref{eq.numevol}), and we plan to experiment
with other functions.  If the IMF is evolving, or is dependent on
environment, this would also affect our results.  The relationship
between star formation rate and radio luminosity
(eq.~\ref{eq.sfrtoLpergal}) might be evolving in addition to its
dependence on an evolving IMF.  Finally, we have not explored the
dependence of our results on cosmological parameters.

This method has the potential to be an important indicator of star
formation history.  Radio luminosity is a reliable indicator of star
formation rate in local galaxies, and is not affected by dust
extinction.  While others are performing similar calculations (Cram
\etal 1998, \cite{cram98a}; Cram 1998, \cite{cram98b}; Mobasher \etal
1999, \cite{mobasher99a}, Serjeant \etal 1998, \cite{serjeant98a}),
the survey data used here are complete to a substantially lower flux
limit, with nearly complete knowledge of optical counterparts and
$\sim$50\% completeness in redshifts.  This allows us to place
stronger constraints on the evolving radio luminosity function and to
probe star formation activity to much higher redshifts.

\section*{Acknowledgments}

We are grateful to Eric Richards for helpful discussions. 
D.H. thanks the National Science Foundation for travel support for
this Symposium.  D.H. and B.P. acknowledge the support of NSF AST
96-16971.

\section*{References}


\begin{thebibliography}{99}

\bibitem{richards98a}
Richards, E.~A., Kellermann, K.~I., Fomalont, E.~B., Windhorst, R.~A.,  \&
  Partridge, R.~B. 1998, AJ, 116, 1039

\bibitem{windhorst95a}
Windhorst, R.~A., Fomalont, E.~B., Kellermann, K.~I., Partridge, R.~B.,
  Richards, E., Franklin, B.~E., Pascarelle, S.~M.,  \& Griffiths, R.~E. 1995,
  Nature, 375, 471

\bibitem{fomalont91a}
Fomalont, E.~B., Kellermann, K.~I., Windhorst, R.~A.,  \& Kristian, J.~A. 1991,
  AJ, 102, 1258

\bibitem{condon92a}
Condon, J.~J. 1992, ARA\&A, 30, 575

\bibitem{cram98a}
Cram, L., Hopkins, A., Mobasher, B.,  \& Rowan-Robinson, M. 1998, ApJ, 507, 155

\bibitem{haarsma98a}
Haarsma, D.~B.,  \& Partridge, R.~B. 1998, ApJ, 503, L5

\bibitem{hauser98a}
Hauser, M.~G.,  et~al. 1998, ApJ, 508, 25

\bibitem{dwek98a}
Dwek, E.,  et~al. 1998, ApJ, 508, 106

\bibitem{helou85a}
Helou, G., Soifer, B.~T.,  \& Rowan-Robinson, M. 1985, ApJ, 298, L7

\bibitem{bridle67a}
Bridle, A.~H. 1967, MNRAS, 136, 219

\bibitem{condon89a}
Condon, J.~J. 1989, ApJ, 338, 13

\bibitem{windhorst93a}
Windhorst, R.~A., Fomalont, E.~B., Partridge, R.~B.,  \& Lowenthal, J.~D. 1993,
  ApJ, 405, 498

\bibitem{partridge97a}
Partridge, R.~B., Richards, E.~A., Fomalont, E.~B., Kellermann, K.~I.,  \&
  Windhorst, R.~A. 1997, ApJ, 483, 38

\bibitem{guiderdoni98a}
Guiderdoni, B., Hivon, E., Bouchet, F.~R.,  \& Maffei, B. 1998, MNRAS, 295, 877

\bibitem{lilly85a}
Lilly, S.~J., Longair, M.~S.,  \& Allington-Smith, J.~R. 1985, MNRAS, 215, 37

\bibitem{katgert88a}
Katgert, P., Oort, M.~J.~A., \& Windhorst, R.~A. 1988, A\&A, 195, 21

\bibitem{martin80a}
Martin, H.~M., Partridge, R.~B., and Rood, R.~T. 1980, ApJ, 240, L79

\bibitem{serjeant98a}
Serjeant, S., Gruppioni, C.,  \& Oliver, S. 1998, preprint astro-ph/9808259

\bibitem{condon84a}
Condon, J.~J. 1984a, ApJ, 284, 44

\bibitem{condon84b}
Condon, J.~J. 1984b, ApJ, 287, 461

\bibitem{elmegreen98a}
Elmegreen, B.~G. 1998, in Unsolved Problems in Stellar Evolution, ed. M.~Livio
  (Cambridge: Cambridge University Press)

\bibitem{recipes2}
Press, W.~H., Flannery, B.~P., Teukolsky, S.~A.,  \& Vetterling, W.~T. 1992,
  Numerical Recipes (2nd ed.) (Cambridge University Press)

\bibitem{pei95a}
Pei, Y.~C.,  \& Fall, S.~M. 1995, ApJ, 454, 69

\bibitem{baugh98a}
Baugh, C.~M., Cole, S., Frenk, C.~S.,  \& Lacey, C.~G. 1998, ApJ, 498, 504

\bibitem{cram98b}
Cram, L. 1998, ApJ, 506, L85

\bibitem{mobasher99a}
Mobasher, B., Cram, L., Georgakakis, A.,  \& Hopkins, A. 1999

\bibitem{hammer95a}
Hammer, F., Crampton, D., Lilly, S.~J., LeFevre, O.,  \& Kenet, T. 1995, MNRAS,
  276, 1085

\bibitem{guiderdoni99a}
Guiderdoni, B., Bouchet, F.~R., Devriendt, J., Hivon, E.,  \& Puget, J.~L.
  1999, preprint astro-ph/9902141


\end{thebibliography}

%


\newpage
 
\end{document}